# Patch-based Contour Prior Image Denoising for Salt and Pepper Noise

*Bo Fu[1], XiaoYang Zhao[1], Yi Li[1], XiangHai Wang[1]*
1. College of Computer and Information Technology, Liaoning Normal University, Dalian, China
E-mail: fubo@lnnu.edu.cn

*Abstract*—The salt and pepper noise brings a significant challenge to image denoising technology, i.e. how to removal the noise clearly and retain the details effectively? In this paper, we propose a patch-based contour prior denoising approach for salt and pepper noise. First, noisy image is cut into patches as basic representation unit, a discrete total variation model is designed to extract contour structures; Second, a weighted Euclidean distance is designed to search the most similar patches, then, corresponding contour stencils are extracted from these similar patches; At the last, we build filter from contour stencils in the framework of regression. Numerical results illustrate that the proposed method is competitive with the state-of-the-art methods in terms of the peak signal-to-noise (PSNR) and visual effects.

## 1. INTRODUCTION

Images are always polluted by noise during image acquisition and transmission because of imperfections in the imaging and capturing process. Removing noise while preserving image details and textures (edge detection, segmentation, etc.) is one of the most important and fundamental issues in image processing. As an important pre-processing step, image denoising algorithms are also widely used in computer vision, pattern recognition and medical image analysis fields. At different stages of imaging, the image may be polluted by corresponding type of noises, for example Poisson noise, Gaussian noise, etc.

Gaussian noise is an additive noise, this kind of noise generally appears during the acquisition of an image. There are huge literature solely dedicated to Gaussian denoising, some of them achieved remarkable results for example famous non-local means(NLM)[1] which exploit the self-similarity property of natural images to perform denoising by aggregation of similar patches. Block-matching and 3D filtering(BM3D)[2] uses 3-D similar patches to perform denoising in transform domain. Another very effective method is the K-SVD algorithm [3], of which the key is how to get optimal dictionary of image patches adapted for the observed noisy data.

Impulse noise is another common type of image noise, it is generally due to data loss happening. Unfortunately, denoising methods for Gaussian noise are totally useless for impulse noise. The main reason is that distributions of two kind of noise are completely different. Two models of impulse noise are generally studied in the literatures, one model is the so-called random-valued impulse noise, which each pixel is replaced with probability p by a random value in the set $\{0, 1, . . . , M\}$, M is set to 255. Another type of image noise is the so-called salt and pepper noise, which each pixel is replaced with a given probability by $[0, \delta)$ or $(255-\delta, 255]$, $\delta$ is generally defined as a very small value like 0,1,2 and 3.

In this paper, we will focus on the second model of impulse noise i.e. salt and pepper noise because it usually bring more obvious visual interference (black or white pixel). For the removal of salt and pepper noise problem, finding noise and repairing it while preserving details and texture information is of great importance. Let **U** as original image, size M*N. Let **V** as observed noisy image, $(i, j)$ be the coordinates of a pixel, $f(i,j)$ is the gray value of pixel (i,j). If pixel (i,j) is noise, the noise model is then defined as:

$$f_{(i,j)} = \begin{cases} N_p & if \quad f \in [0...T_p) \\ N_s & if \quad t \in (T_s, 255] \\ V_{(i,j)} & if \quad t \in (T_p, T_s) \end{cases} \quad (1)$$

In the past decade, a number of denoising algorithms have been proposed. One well-known algorithm family is nonlinear filters such as median filters [4,5]. These kinds of methods adopt the median value of the pixel value in the local area as the basis for restoration. Switching filter and its extension introduce a switching templet to decrease noise's interference during repairing step[6-12]. Though some achievements have been achieved, these methods mentioned above only take local region into account in analysis and repair stage. Too little information leads to inaccurate identify noise and instable repair work.

Buades firstly proposed non local concept, and design a non-local means for Gaussian noise[1]. Based non-local idea, famous BM3D was born [2,13], it constructs non local collaborative filtering in frequency domain. Drawing on this idea, some denoising algorithms for salt and pepper have been proposed forward. Nasri et al. improved classical non-local means by introducing a switching filter to reduce the interference of impulse noise [14], Varghese J proposed an adaptive version[15]. All methods mentioned above have two common points. First, they are all patches based methods. Second, the computation of similarity takes place in a non-local scope. The above characteristic brings both advantages and challenges. Patch contain more information(texture, edge, etc.) than single pixel, but it is more difficult to estimate and mine effective information under high noise density condition. A large number of patches will pose a new challenge in precision measurement between patches especially these patches are disturbed by noise. In view of these problems, scholars in the field of CV have launched some attempts. Delon designed a statistical estimator to estimate patch and compute distance between patches [16]. Meanwhile, some clusters that consist of a few patches are abandoned to suppress noise[17-20]. Some low-rank methods and patch retrieval methods are also be proposed[21-24,30-33]. But they generally did not dig the texture information of the patch meanwhile restrain the noise.

Given the above problems, the idea of this paper is relies on two contributions. First, we introduce Total Variation(TV) to describe sharp trend of texture in each patch. Second, for each patch we will generate a geometric structure according patches' texture, and build regression restoration in non-local range. The rest of this paper is organized as follows: In section 2, we define a basic form of TV model, and compute a direction template to approximate function meanwhile resisting noise. In section 3, using generated a geometric structure in section 2, we build a Non local filtering for patches regression. Section 4 shows numerous experimental results and comparisons with several related algorithms on several types of nature images.

## 2. PATCH CONTOURS ESTIMATION BASED ON TOTAL VARIATION

TV model is proposed by Rudin firstly [25], a significant advantage of TV model is that it allows discontinuity points exist in variational function space BV($\Omega$) (it leads to an underlying sparse solution). Set $\Omega \subset \mathbf{R}^2$, $U \in L^1(\Omega)$ is a bounded open collection, it is usually assumed to be a Lipschitz domain. Assume **U** can be expressed as function $u(i,j)$, and u is smooth. The TV express of $u$ is:

$$\min_{u \in BV(\Omega)} TV[u] = \int_\Omega |\nabla u| dxdy, \quad \nabla u = (u_x, u_y) \quad (2)$$

Here, $u$ satisfies the following constraint conditions:

$$\int_{\Omega} u(x,y)dxdy = \int_{\Omega} u_0(x,y)dxdy, \quad \frac{1}{|\Omega|}\int_{\Omega}(u(x,y)-u_0(x,y))^2 dxdy = \sigma^2 \tag{3}$$

Further, the document [26] defines 8 kinds of contour templates to discretize the contour stencils(CS) information of the integral region, CS is defined as:

$$(S^*[u])(k) := \sum_{m,n \in \mathbf{R}^2} S(m,n)|u_{k+m} - u_{k+n}| \tag{4}$$

For each pixel, they find aimed contour stencils which it's $(S^*[u])(k)$ is minimum. But, these templates only cover 8 directions and they are under the assumption that the function is smooth.

So in our work, we firstly choose median value to change salt and pepper noise to avoid noise interference before generating contour stencils. And then we extend 8 directions to 3 types of directional templates including horizontal direction, vertical direction, and diagonal direction. Each direction including 8 contour stencils and corresponding discretization templates. The specific definitions are given in Table 1.

Table 1 Multi-angle template with 3 directions

| Horizontal | $S_1^{(1)}$ | $S_2^{(1)}$ | $S_3^{(1)}$ | $S_4^{(1)}$ | $S_5^{(1)}$ | $S_6^{(1)}$ | $S_7^{(1)}$ | $S_8^{(1)}$ |
|---|---|---|---|---|---|---|---|---|
| Direction | (diagram) | (diagram) | (diagram) | (diagram) | (diagram) | (diagram) | (diagram) | (diagram) |
| Angle | $\pi - \arctan\frac{1}{2}$ | $\arctan\frac{1}{2}$ | $\frac{1}{8}\pi$ | $\frac{7}{8}\pi$ | $-\pi + \arctan\frac{1}{2}$ | $-\arctan\frac{1}{2}$ | $-\frac{1}{8}\pi$ | $-\frac{7}{8}\pi$ |
| Discretization matrix | $\begin{bmatrix}1&0&0\\0&-2&1\\0&0&0\end{bmatrix}$ | $\begin{bmatrix}0&0&1\\1&-2&0\\0&0&0\end{bmatrix}$ | $\begin{bmatrix}0&0&1\\0&-2&1\\0&0&0\end{bmatrix}$ | $\begin{bmatrix}1&0&0\\1&-2&0\\0&0&0\end{bmatrix}$ | $\begin{bmatrix}0&0&0\\0&-2&1\\1&0&0\end{bmatrix}$ | $\begin{bmatrix}0&0&0\\1&-2&0\\0&0&1\end{bmatrix}$ | $\begin{bmatrix}0&0&0\\0&-2&1\\0&0&1\end{bmatrix}$ | $\begin{bmatrix}0&0&0\\1&-2&0\\1&0&0\end{bmatrix}$ |
| vertical | $S_1^{(2)}$ | $S_2^{(2)}$ | $S_3^{(2)}$ | $S_4^{(2)}$ | $S_5^{(2)}$ | $S_6^{(2)}$ | $S_7^{(2)}$ | $S_8^{(2)}$ |
| Direction | (diagram) | (diagram) | (diagram) | (diagram) | (diagram) | (diagram) | (diagram) | (diagram) |
| Angle | $-\pi + \arctan 2$ | $-\arctan 2$ | $\frac{3}{8}\pi$ | $\frac{5}{8}\pi$ | $\pi - \arctan 2$ | $\arctan 2$ | $-\frac{3}{8}\pi$ | $-\frac{5}{8}\pi$ |
| Discretization matrix | $\begin{bmatrix}0&1&0\\0&-2&0\\1&0&0\end{bmatrix}$ | $\begin{bmatrix}0&1&0\\0&-2&0\\0&0&1\end{bmatrix}$ | $\begin{bmatrix}0&1&1\\0&-2&0\\0&0&0\end{bmatrix}$ | $\begin{bmatrix}1&1&0\\0&-2&0\\0&0&0\end{bmatrix}$ | $\begin{bmatrix}1&0&0\\0&-2&0\\0&1&0\end{bmatrix}$ | $\begin{bmatrix}0&0&1\\0&-2&0\\0&1&0\end{bmatrix}$ | $\begin{bmatrix}0&0&0\\0&-2&0\\0&1&1\end{bmatrix}$ | $\begin{bmatrix}0&0&0\\0&-2&0\\1&1&0\end{bmatrix}$ |
| diagonal | $S_1^{(3)}$ | $S_2^{(3)}$ | $S_3^{(3)}$ | $S_4^{(3)}$ | $S_5^{(3)}$ | $S_6^{(3)}$ | $S_7^{(3)}$ | $S_8^{(3)}$ |
| Direction | (diagram) | (diagram) | (diagram) | (diagram) | (diagram) | (diagram) | (diagram) | (diagram) |
| Angle | $\frac{6}{8}\pi$ | $\frac{2}{8}\pi$ | $-\frac{6}{8}\pi$ | $-\frac{2}{8}\pi$ | $\frac{4}{8}\pi$ | $\frac{8}{8}\pi$ | $\frac{0}{8}\pi$ | $-\frac{4}{8}\pi$ |

| Discretization matrix | $\begin{bmatrix} 0 & 1 & 0 \\ 1 & -2 & 0 \\ 0 & 0 & 0 \end{bmatrix}$ $\begin{bmatrix} 0 & 1 & 0 \\ 0 & -2 & 1 \\ 0 & 0 & 0 \end{bmatrix}$ $\begin{bmatrix} 0 & 0 & 0 \\ 1 & -2 & 0 \\ 0 & 1 & 0 \end{bmatrix}$ $\begin{bmatrix} 0 & 0 & 0 \\ 0 & -2 & 1 \\ 0 & 1 & 0 \end{bmatrix}$ $\begin{bmatrix} 1 & 0 & 1 \\ 0 & -2 & 0 \\ 0 & 0 & 0 \end{bmatrix}$ $\begin{bmatrix} 1 & 0 & 0 \\ 0 & -2 & 0 \\ 1 & 0 & 0 \end{bmatrix}$ $\begin{bmatrix} 0 & 0 & 1 \\ 0 & -2 & 0 \\ 0 & 0 & 1 \end{bmatrix}$ $\begin{bmatrix} 0 & 0 & 0 \\ 0 & -2 & 0 \\ 1 & 0 & 1 \end{bmatrix}$ |
|---|---|

We denotes $S_{k_d}^{(d)}$ as our extended contour templates, and upper corner mark $d$ represents 3 kinds of directional classes(i=1, 2, 3), under corner mark $k$ represents 8 directional contour templates(k=1,2,…8). Where $(S_{k_d}^{(d)}.*[u_{(i,j)}])$ is seen a discrete estimate value i.e. $(S_{k_i}^{(d)}.*[u_{(i,j)}]) \approx \|u_{k_i}^d\|_{TV(C^{(d)})}$. We can generate the corresponding contour stencil by minimum value of $(S_{k_d}^{(d)}.*[u_{(i,j)}])$. For a patch p, we denote corresponding contour stencils as **cs**.

## 3. PROPOSED METHOD

We introduce the proposed patch-based contour prior approach denoising algorithm for salt and pepper noise. In this approach, we firstly design a step of generating contour templets pixel by pixel. For each patch, there is a corresponding contour structure. Second, we use nearest-neighbors (NN) to find similar patches. Then, our algorithm use a regression method to repair central noise in patches. The basic flow of our algorithm is as follows:

---
**Patch-based contour prior denoising algorithm**

input : noisy image V,
output: denoised image V',
cut V into M*N patches
**for** q from 1 to M*N **do**
   if the central pixel of patch $P_q$ is noise;
   **for** q1 form 1 to 28 **do**
      Compute the Equ.4;
      Return the minim value;
   Generate the contour stencils $cs_q$.
   **for** q2 form 1 to M*N
      use the nearest-neighbors to find *mm* the most similar contour stencils;
      use regression method to repair the central noise in $P_q$
Return V'

---

### 3.1 weighted Euclidean distance

In order to restore noise in the target contour structure, we need search for enough similar contour stencils. It is well known that $l^2$ distance is good measurement method between patches. But in this stance, all patches generally contain large number of salt and pepper noise(maximum and minimum value), these noise will cause great interference to the process of measurement. In order to avoid this problem, we use a probabilistic approach to compute distance avoid noise interference. The specific way is as follows:

For a target patch P, we firstly compute P's similarities with other patches Q in the whole noisy image scope. Suppose that the target P and Q contain $n$ pixels, calculate the distance between two patches pixel by pixel, and then summing up these distances with a weight $w$. The specific weights of w are generated by the use 0-1 distribution. We can estimate roughly estimate the number of noise $n_n$ with classical identifier, and get the probability *p* of 0-1 distribution by $n_n/n$. Denotes the distance of each pair of points as $D_p(P_{kk},Q_{kk})$, kk=1,2,..n. We sorted the distances of $D_p(P_{kk},Q_{kk})$ from large to small. Obviously, the greater the distance, the more likely it is noise. So, the first $n_n$ weithts's value in the sequence are 0, and the others are $1/(n- n_n)$. Then we can compute the weighted Euclidean distance as follows:

$$D(P,Q) = \sum_{kk=1}^{n} w_{kk} |P_{kk} - Q_{kk}|^2 \tag{5}$$

According to the distance $D(P,Q)$, we can use nearest-neighbors methods to find a set of similar patches **R** in the global scope of noisy image.

## 3.2 Contour prior filter

For a corrupted pixel (i,j) in patch P, we can get a set of most similar patches Q. We denotes the contour stencil of P as $CS_p$, the contour stencil of Q as $CS_q$. We adopt contour stencils to build filter because they contains more priori information. In the regression framework, we can use these contour stencils to construct non local filtering. Let the restored gray value of the corrupted pixel be $\tilde{f}_{(i,j)}$, which is obtained by weighted averaging as follows:

$$\tilde{f}_{(kk)} = \sum_{\forall (q) \in R} w_{(CSp,CSq)} \cdot f_{(kk)} \tag{6}$$

Here, any CS*q* belong to the similar patches **R**, $w_{(CSp,CSq)}$ is the weight according to the contour stencil $CS_p$ and $CS_q$. Weight $w_{(CSp,CSq)}$ is calculated using the similarity $s_{(CSp,CSq)}$ between each reference contour stencil and the target contour stencil, the corresponding formula is as follows:

$$w_{(CSp,CSq)} = \frac{s_{(CSp,CSq)}}{\sum_{1}^{mm} s_{(CSp,CSq)}} \tag{7}$$

Here, mm is the number of similar contour stencils. The specific definition of $s_{(CSp,CSq)}$ is as follow:

$$s_{(CSp,CSq)} = \frac{1}{e^{\frac{\|CSp-CSq\|^2}{\sigma}} \ln \sigma} \tag{8}$$

Here, $\sigma$ is used to control strength of the filtering corrosion.

## 4. EXPERIMENTS

### 4.1 Experimental Setting

***Dataset*** We choose some classical nature images as experimental dataset, it contains 11 commonly used images for denoising, e.g., "Lena", "Baboon" and "Pepper", "Bark" etc.

***Baseline algorithm*** We use four existing denoising algorithms as baselines, including adaptive median filter (AMF), decision based algorithm (DBA[6]), method based on pixel density filter(BPDF)[27], Decision based Unsymmetrical Trimmed Variants (DBUTVF)[28], Adaptive Center Weighted Median Filter(ACWMF)[32], and Patch-based Approach to Remove Impulse-Gaussian Noise(PARIGI[16]). For fair comparison, all methods' code implementations are their publicly available versions and their parameters are set following the guidelines in original articles. The peak signal-to-noise ratio (PSNR) is adopted to measure the objective performance of our algorithm.

***Evaluation Metric*** The peak signal-to-noise ratio (PSNR) is adopted to measure the objective performance of our algorithm. PSNR is defined as follows.

$$\text{PSNR} = 10\log_{10}(\frac{255^2}{MSE}) \tag{9}$$

## 4.2 Performance Comparison

Experiments were carried out on corrupted images with various noise densities (10%–90%). The denoising results of Image Lena are compared in Table 2.

Table 2 PSNR scores result

| Level | AMF | DBA | BPDF | ACWMF | DBUTVF | PARIGI | Our algorithm |
|---|---|---|---|---|---|---|---|
| 10% | 38.61 | 41.50 | 40.04 | 39.19 | 41.50 | 37.65 | **47.48** |
| 20% | 36.10 | 37.47 | 35.84 | 30.99 | 37.48 | 35.83 | **42.74** |
| 30% | 33.79 | 34.81 | 33.09 | 24.08 | 34.85 | 33.93 | **39.15** |
| 40% | 32.22 | 32.30 | 30.59 | 19.06 | 34.23 | 32.32 | **36.39** |
| 50% | 30.39 | 30.11 | 28.13 | 15.28 | 30.14 | 29.91 | **32.94** |
| 60% | 28.81 | 28.08 | 25.89 | 12.27 | 28.26 | 27.45 | **29.67** |
| 70% | **27.11** | 25.58 | 23.16 | 10.01 | 25.94 | 25.47 | 26.14 |
| 80% | **24.89** | 23.22 | 17.68 | 8.11 | 23.36 | 23.18 | 23.16 |
| 90% | 20.35 | 19.91 | 10.85 | 6.64 | **20.63** | 16.86 | 16.56 |

It can be seen that our algorithm obtain great superiority in low noise density condition. It not obtain the best scores in high noise density condition because TV model no longer works.

At the same time, visual comparisons are also provided in Figure 2, some details are enlarged in lower right corner of each picture.

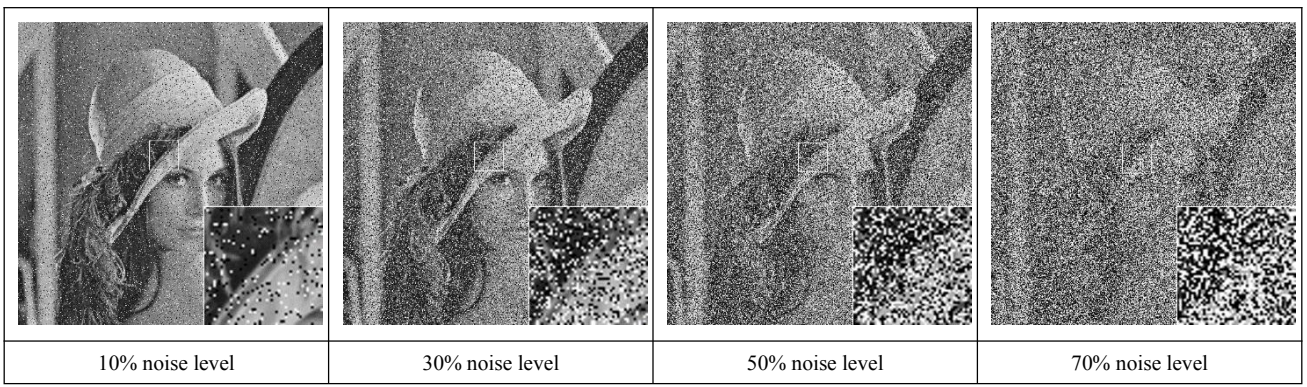

| 10% noise level | 30% noise level | 50% noise level | 70% noise level |

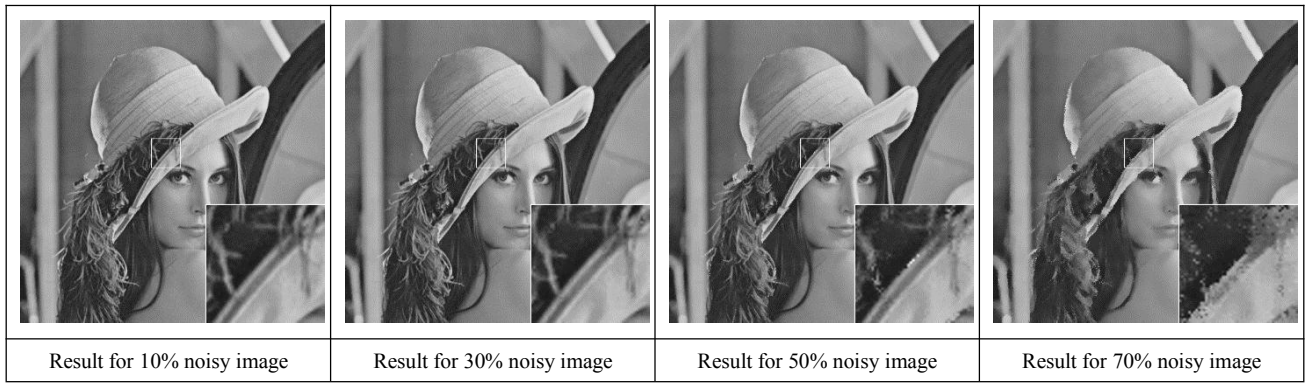

| Result for 10% noisy image | Result for 30% noisy image | Result for 50% noisy image | Result for 70% noisy image |

Fig. 2 Visual quality results of Image Lena

It can be seen that our algorithm obtains good visual effect. Due to TV model, our algorithm can protect edge information better.

## 5. CONCLUSION

In this paper, an image denoising algorithm based on patch-based contour prior for salt and pepper noise is proposed. First, a discrete total variation model is introduced to extract contour structures. Second, a weighted Euclidean distance is designed to search the most similar patches, then, corresponding contour stencils are extracted from these similar patches; At the last, we build filter from contour stencils in the framework of regression. Numerical results illustrate that the proposed method is competitive with the state-of-the-art methods in terms of the peak signal-to-noise (PSNR) and visual effects.

## ACKNOWLEDGEMENTS

This work is supported by the National Natural Science Foundation of China (NSFC) Grant No. 61702246, 41671439, Liaoning Province of China General Project of Scientific Research No. L2015285, Doctoral Start-up Foundation of Liaoning Province No. 201601243.